\documentclass[aps,prd,onecolumn,groupedaddress,showpacs,nofootinbib,amssymb]{revtex4}
\usepackage{graphicx,bm,color}
\usepackage{amsmath}
\usepackage{amssymb}
\usepackage{amsfonts}
\newcommand{\be}{\begin{equation}}
\newcommand{\ee}{\end{equation}}
\newcommand{\bea}{\begin{eqnarray}}
\newcommand{\eea}{\end{eqnarray}}
\newcommand{\beaa}{\begin{eqnarray*}}
\newcommand{\eeaa}{\end{eqnarray*}}

\newcommand{\nn}{\nonumber \\}
\newcommand{\e}{\mathrm{e}}

\allowdisplaybreaks[4]

\begin{document}
\tolerance=5000

\title{$F(R)$ Gravity with an Axion-like Particle: Dynamics, Gravity Waves, Late and Early-time Phenomenology}
\author{Shin'ichi~Nojiri,$^{1,2}$\,\thanks{nojiri@gravity.phys.nagoya-u.ac.jp}
S.~D.~Odintsov,$^{3,4}$\,\thanks{odintsov@ieec.uab.es}
V.~K.~Oikonomou,$^{5,6}$\,\thanks{v.k.oikonomou1979@gmail.com} }
\affiliation{ $^{1)}$ Department of Physics, Nagoya University,
Nagoya 464-8602, Japan \\
$^{2)}$ Kobayashi-Maskawa Institute for the Origin of Particles
and the Universe, Nagoya University, Nagoya 464-8602, Japan \\
$^{3)}$ ICREA, Passeig Luis Companys, 23, 08010 Barcelona, Spain\\
$^{4)}$ Institute of Space Sciences (IEEC-CSIC) C. Can Magrans
s/n,
08193 Barcelona, Spain\\
$^{5)}$ Department of Physics, Aristotle University of
Thessaloniki, Thessaloniki 54124,
Greece\\
$^{6)}$ International Laboratory for Theoretical Cosmology, Tomsk
State University of Control Systems and Radioelectronics (TUSUR),
634050 Tomsk, Russia}

\tolerance=5000

\begin{abstract}
In this work we investigate several theoretical and
phenomenological implications of a scalar -$F(R)$ gravity
containing a non-minimal coupling to the scalar curvature. This
kind of model is a generalization of axion-$F(R)$ gravity models,
so we shall examine several implications of the latter theory.
Firstly we study in detail the Einstein frame picture of the
model, and also we discuss the dynamics of the cosmological
system. By appropriately using the equations of motion, we
demonstrate that an arbitrary cosmological evolution can be
realized. Also we study the gravitational waves of the theory, and
we demonstrate that the speed of their propagation is the same as
in $F(R)$ gravity, but there is the possibility of enhancement or
dissipation of the gravitational waves, an effect quite similar to
the propagation of gravity waves in a viscous fluid. Finally, we
examine the energy momentum tensor and we investigate which
quantities related to it are conserved. We also present the
constraints imposed by the radiation domination era on the
non-minimal coupling of the axion scalar field to the scalar
curvature.
\end{abstract}


\maketitle

\section{Introduction}

Dark matter is one of the persisting problems in modern
theoretical particle physics and cosmology, and up to date, no
experimental verification occurred. Observational data coming from
galactic scale structures, like the bullet cluster, or from
galactic rotation curves, strongly indicate that dark matter seems
to control the dynamics of galaxies rotation and also seems to
control the collision of galaxies. In addition, the observation of
the bullet cluster, strongly indicates that dark matter is a
particle, and halos of dark matter particles accompany the visible
structure of a galaxy. Although, the dark matter effects can be
mimicked by modified gravity itself
\cite{reviews4,reviews5,reviews6,reviews1,reviews2,reviews3}, see
also  \cite{Capozziello:2006ph,Nojiri:2016vhu}, observations seem
to favor the particle nature of dark matter for the moment at
least. Up to present date, the dark matter searches utterly failed
to identify any dark matter particle, and to our opinion this is
because most dark matter searches focused on mass scales for the
weakly interacting massive particle (WIMP) of the order up to GeV
or even hundred GeV \cite{Oikonomou:2006mh}. However, the
scientific community searching for WIMPs now focuses on mass
scales of eV or much more smaller mass scales. One of the most
promising class of WIMPs with tiny mass is the axion, or any axion
like particle predicted from low scale compactifications of string
theory
\cite{Marsh:2015xka,Marsh:2017yvc,Odintsov:2019mlf,Cicoli:2019ulk,Fukunaga:2019unq,Caputo:2019joi,maxim,Auriol:2018ovo,Ioannisian:2017srr}.
In fact several experimental and observational proposals exist
already in the literature
\cite{Du:2018uak,Henning:2018ogd,Ouellet:2018beu,Safdi:2018oeu,Rozner:2019gba,Avignone:2018zpw,Caputo:2018vmy,Caputo:2018ljp,Millar:2016cjp,Majorovits:2017ppy,Anastassopoulos:2017ftl,TheMADMAXWorkingGroup:2016hpc},
see also \cite{Lawson:2019brd}, and many researchers aim to
identify in the laboratory this elusive particle. To our opinion,
if the axion exists, it will be found in the next 10-15 years due
to extensive experimental searches now performed, mainly based on
the fact that axions and photons interact in the presence of
magnetic fields
\cite{Balakin:2009rg,Balakin:2012up,Balakin:2014oya}, and the
axion is the last resort of particle dark matter, unless nature
hides the WIMPs to a supersymmetry breaking scale, and this will
be another surprise for theorists. The axion is known to provide a
very good candidate for dark matter, or at least some of the dark
matter existing in the Universe \cite{Marsh:2015xka}, and when we
refer to the axion, this should not be confused with the QCD
axion, which is a pseudo-scalar field, but we refer to any
axion-like particle, with a primordial $U(1)$ broken symmetry.
Particularly, these are known as misalignment axion fields, which
are canonical scalar fields. In Ref.~\cite{Odintsov:2019evb}, we
investigated an effective model of $F(R)$ gravity in the presence
of an misalignment axion field, and by using the axion field
dynamics known from the literature \cite{Marsh:2015xka}, we
demonstrated that the axion via a non-minimal coupling to the
scalar curvature affects the late-time era, making possible for
the dark energy era to be realized. Some striking features of the
model were the facts that firstly, the early-time era was
controlled by the well-known $R^2$ gravity
\cite{Starobinsky:1982ee}, and the axion field after it starts
oscillating, for cosmic times for which $m_a\geq H$, so during and
after the reheating era, its energy density scales as $\rho_a\sim
a^{-3}$. Effectively it describes a dark matter condensate field,
which on average yields an effective equation of state (EoS)
parameter $\langle w \rangle =0$.

In this work, we shall further explore several theoretical and
phenomenological features of the model \cite{Odintsov:2019evb},
and of some extensions of this model, to include non-canonical
scalar fields. So the study is focused on theoretical problems and
implications of $F(R)$ gravity in the presence of a non-canonical
in general scalar field, and we emphasize on the phenomenology of
the theory containing the canonical scalar field, which is a
subcase. Particularly, we shall consider the Einstein frame
picture of the scalar-$F(R)$ gravity, and we shall derive the
no-ghost conditions. Also we shall consider the general dynamics
of the model and we shall demonstrate that the resulting
gravitational equations of motion can be used as a reconstruction
technique, which enables us to realize an arbitrary cosmological
evolution. In addition, we shall study the gravitational wave
spectrum, and we shall find several constraints on the
scalar-$F(R)$ gravity theory. Finally, the energy conditions are
considered, and by using these, for the radiation domination era,
we evince how the scalar-$F(R)$ gravity theory can be constrained,
and specifically how the non-minimal coupling of the axion to the
curvature can be constrained.

The motivation for using extensions of $F(R)$ gravity including
scalar fields is threefold. Firstly, $F(R)$ gravity has been
considered and actively investigated as a model quite elegantly
describing dark energy \cite{Capozziello:2003gx,Nojiri:2003ft}. In
the $F(R)$ gravity, appears a scalar particle called scalaron
which interacts with the standard model particles very weakly and
therefore the lifetime of this particle could be rather long. Then
some scenarios where the scalaron can be dark matter particle have
been also considered
\cite{Nojiri:2008nt,Cembranos:2008gj,Katsuragawa:2016yir}. This
scenario is very fascinating because the $F(R)$ gravity might
explain both of the dark energy and dark matter. Since the
scalaron is directly interacting with the standard model
particles, however, the scalaron can decay into the standard
particles even if the interaction is very weak and we need to
consider some scenario to suppress this decay. One easy way to
avoid this problem is to introduce another scalar field and we may
regard the particle corresponding to the field as a dark matter.
Secondly, the Starobinsky model  seems to be quite robust against
the Planck 2018 \cite{Akrami:2018odb} constraints on inflation,
and thirdly, after the Higgs particle discovery
\cite{Aad:2012tfa}, the scalar fields seem to be an inherent
constituent of the primordial era. Thus in this work we shall
further explore the theoretical and phenomenological implications
of the misalignment axion-$F(R)$ gravity.

This paper is organized as follows: In section II, we briefly
recall the essential features of the misalignment axion-$F(R)$
gravity model developed in Ref.~\cite{Odintsov:2019evb}. In
section III, we study the Einstein frame picture of the model, and
we derive the no-ghost criteria, while in section IV, we discuss
the dynamics of the model. We also provide a general
reconstruction technique that can be used to realize any arbitrary
cosmological evolution. In section V we study the gravitational
waves of the axion-$F(R)$ gravity model, and we consider several
constraints related with the observations. In section VI, we
discuss the energy conditions, and we study the dynamics of the
model for several eras, and we demonstrate how the radiation
domination era may impose a constraint on the non-minimal coupling
of the axion to the scalar curvature. Finally, the conclusions
follow at the end of the paper.

\section{Overview of the non-minimally Coupled Axion-$F(R)$ Gravity Model\label{seca}}

In this section we shall briefly recall the main outcomes of the
model developed in Ref.~\cite{Odintsov:2019evb}, in order to put
the present paper in proper context, and to have a reference point
for the studies that follow. The axion-like $F(R)$ gravity model
of Ref.~\cite{Odintsov:2019evb} was based on the following
gravitational action,
\begin{equation}
\label{mainaction} \mathcal{S}=\int d^4x\sqrt{-g}\left[
\frac{1}{2\kappa^2}F(R)+\frac{1}{2\kappa^2}h(\phi)G(R)-\frac{1}{2}\partial^{\mu}\phi\partial_{\mu}\phi-V(\phi)
\right]\, ,
\end{equation}
with $\kappa^2=\frac{1}{8\pi G}$, and $G$ stands for Newton's
gravitational constant. The $F(R)$ gravity chosen in Ref.~\cite{Odintsov:2019evb}
was the well-known $R^2$ gravity,
\begin{equation}\label{starobinsky}
F(R)=R+\frac{1}{36H_i^2}R^2\, .
\end{equation}
Also the coupling $h(\phi)$ to $G(R)$, was assumed to be,
\begin{equation}\label{hphichoice}
h(\phi)\sim\frac{1}{\phi^{\delta}}\, ,
\end{equation}
choosing $\delta>0$, and also the $G(R)$ function is,
\begin{equation}\label{GRfunction}
G(R)\sim R^{\gamma}\, ,
\end{equation}
with the parameter $\gamma$ belonging to the interval $0<\gamma
<0.75$. The model of Ref.~\cite{Odintsov:2019evb} is an effective
model which relies mostly on the axion field dynamics. The axion
field at early times, when the axion mass $m_a$ satisfies $m_a\ll
H$, is considered frozen in its vacuum expectation value $\phi_i$,
with the potential being,
\begin{equation}\label{axionpotential}
V(\phi(t))\simeq \frac{1}{2}m_a^2\phi^2_i(t)\, .
\end{equation}
The following initial conditions at early times are assumed,
\begin{equation}\label{axioninitialconditions}
\dot{\phi}(t_i)=\zeta \ll 1,\,\,\,\phi(t_i)=f_a\theta_a\, ,
\end{equation}
with $t_i$ being the cosmic time corresponding to the inflationary
era, and $f_a$ being the axion decay constant. In view of the
initial conditions, the axion is considered frozen, thus it
contributes a minor overdamped cosmological constant during
inflation, therefore the evolution during the inflationary era is
governed solely by the $R^2$ gravity. As the Universe expands and
cools, when $m_a\geq H$, the axion field starts to oscillate. For
all cosmic times, the axion field equation of motion is,
\begin{equation}\label{scalarfieldeqnduringradiation}
\ddot{\phi}+3H\dot{\phi}+m_a^2\phi=0\, ,
\end{equation}
so for $m_a\geq H$, by assuming a slow-varying oscillatory
behavior for the axion, of the form,
\begin{equation}\label{solutionaxionradandaft}
\phi (t)= A(t)\cos (m_a t)\, ,
\end{equation}
with $A(t)$ quantifying the slow-varying behavior, since it
satisfies,
\begin{equation}\label{Atsolutionconstraints}
\frac{\dot{A}}{m_a}\sim \frac{H}{m_a}\sim \epsilon\ll 1\, ,
\end{equation}
by solving the equation of motion~(\ref{scalarfieldeqnduringradiation})
with the ansatz~(\ref{solutionaxionradandaft}), we get the solution,
\begin{equation}\label{solutionforA}
A\sim a^{-3/2}\, .
\end{equation}
Hence, for cosmic times after the inflationary era, during the
reheating and until late times, the scalar field behaves as,
\begin{equation}\label{solutionaxionradandaftpaper2}
\phi (t)= a^{-3/2}\cos (m_a t)\, .
\end{equation}
Thus, at early times, the axion field has an effective equation of
state (EoS) parameter $w_a=-1$, while for cosmic times $m_a\gg H$,
it has an average EoS of the form $w_a=0$, which describes dark
matter. Finally, as it was shown in Ref.~\cite{Odintsov:2019evb},
the late-time dynamics of the model is governed solely by the term
$\sim h(\phi) G(R)$, with $h(\phi)\sim \phi^{-\delta}$ and
$G(R)\sim R^{\gamma}$ and $0<\gamma <0.75$.

In the following sections we shall discuss several issues related
with the phenomenological aspects of the axion-$F(R)$ gravity
model. Particularly, we shall consider the Einstein frame
counterpart theory, and we discuss the constraints imposed by the
non-ghost condition, and several other theoretical issues related
to the Einstein frame picture. Also we shall present the
reconstruction techniques for the general scalar-$F(R)$ gravity,
which enables us to realize any cosmological solution, given the
Hubble rate. This is particularly useful for cosmological eras in
between the inflationary and dark energy era. Also we shall
discuss the gravitational waves solutions corresponding to early
times, and finally, we shall examine if the energy conditions are
satisfied, and how the radiation domination era may constrain the
coupling of the scalar field $h(\phi)$ to the $G(R)$ gravity term.

\section{The Einstein Frame Picture: Constraints and Limitations \label{Sec1}}

Let us firstly consider the Einstein frame picture of the
scalar-$F(R)$ gravity theory, and we discuss in some detail some
theoretical implications and constraints of the model.

Let us consider a more general gravitational action in comparison
to the one appearing in Eq.~(\ref{mainaction}), of the form,
\begin{equation}
\label{FhGV1} S = \int d^4 x \sqrt{-g} \left[ \frac{1}{2\kappa^2}
\left\{ F(R) + h(\phi) G(R)\right\}
 - \frac{1}{2} \omega (\phi ) \partial_\mu \phi \partial^\mu \phi - V (\phi)
+ \mathcal{L}_\mathrm{matter} \left( g_{\mu\nu}, \Phi_i \right)
\right] \, ,
\end{equation}
where $F(R)$ and $G(R)$ are functions of the scalar curvature $R$
and $h(\phi)$, $\omega(\phi)$, and $V(\phi)$ are functions of the
scalar field $\phi$. Also, $\mathcal{L}_\mathrm{matter} \left(
g_{\mu\nu}, \Phi_i \right) $ is the Lagrangian density of the
matter fluids present -if any- and $\Phi_i$'s express the matter
fields. We now rewrite the action~(\ref{FhGV1}) by introducing the
auxiliary fields $A$ and $B$ as follows,
\begin{equation}
\label{FhGV2}
S_{AB} = \int d^4 x \sqrt{-g} \left[ \frac{1}{2\kappa^2} \left\{ B \left( R - A \right)
+ F(A) + h(\phi) G(A)
 - \frac{1}{2} \omega (\phi ) \partial_\mu \phi \partial^\mu \phi - V (\phi) \right\}
+ \mathcal{L}_\mathrm{matter} \left( g_{\mu\nu}, \Phi_i \right) \right] \, .
\end{equation}
Then by varying the action with respect to the auxiliary scalar
$A$, we obtain,
\begin{equation}
\label{FhGV3}
B = F'(A) + h(\phi) G'(A)\, .
\end{equation}
Then the condition for the absence of the anti-gravity or the
condition that the graviton is not ghost is given by,
\begin{equation}
\label{FhGV4}
F'(A) + h(\phi) G'(A) = F'(R) + h(\phi) G'(R) > 0 \, .
\end{equation}
Obviously, this condition is satisfied by the gravitational action
(\ref{mainaction}). Upon redefining the scalar field $B$ by using
a new scalar field $\sigma$ as $B=\e^{\sigma}$, we assume that
Eq.~(\ref{FhGV3}) can be solved with respect to $A$ as $A=A(\phi,
\sigma)$. Then the action~(\ref{FhGV2}) can be rewritten as
follows,
\begin{align}
\label{FhGV5}
S_{\sigma\phi} = \int d^4 x \sqrt{-g} & \left[ \frac{1}{2\kappa^2} \left\{ \e^\sigma
\left( R - A\left( \phi , \sigma\right) \right)
+ F\left(A \left( \phi , \sigma\right) \right)
+ h(\phi) G \left( A\left( \phi , \sigma\right) \right)
 - \frac{1}{2} \omega (\phi ) \partial_\mu \phi \partial^\mu \phi - V (\phi) \right\}
\right. \nn
& + \mathcal{L}_\mathrm{matter} \left( g_{\mu\nu}, \Phi_i \right) \Bigr] \, .
\end{align}
By the scale transformation of the metric,
\begin{equation}
\label{FhGV6}
g_{\mu\nu} = \e^{-\sigma} {\tilde g}_{\mu\nu} \, ,
\end{equation}
the action~(\ref{FhGV5}) can be rewritten in the Einstein frame,
and it is equal to,
\begin{align}
\label{FhGV7}
S_\mathrm{E} =& \int d^4 x \sqrt{-\tilde g} \left[ \frac{1}{2\kappa^2} \left\{
\tilde R - \frac{3}{2} \partial_\mu \sigma \partial^\mu \sigma
 - \frac{1}{2} \e^{-\sigma} \omega (\phi ) \partial_\mu \phi \partial^\mu \phi
 - U \left( \phi, \sigma \right) \right\}
+ \e^{-2\sigma} \mathcal{L}_\mathrm{matter}
\left( \e^{-\sigma} {\tilde g}_{\mu\nu}, \Phi_i \right) \right] \, , \nn
U \left( \phi, \sigma \right) \equiv &
\e^{-\sigma} A\left( \phi , \sigma\right)
+ \e^{-2\sigma} \left\{ - F\left(A \left( \phi , \sigma\right) \right)
 - h(\phi) G \left( A\left( \phi , \sigma\right) \right)
+ V (\phi) \right\} \, .
\end{align}
In addition to Eq.~(\ref{FhGV4}), the condition that ensures the
absence of ghost degrees of freedom is,
\begin{equation}
\label{FhGV8} \omega (\phi ) > 0 \, ,
\end{equation}
which is clearly satisfied by the gravitational action~(\ref{mainaction}).
Also, let us quote here a marginal remark,
related to the swampland criteria, which is the following, the
constraint given by the so-called swampland conjecture could have
the following form,
\begin{equation}
\label{FhGV9}
\sqrt{ \frac{1}{3} \left( \frac{\partial U}{\partial \sigma} \right)^2
+ \frac{\e^\sigma}{\omega(\phi)} \left( \frac{\partial U}{\partial \phi} \right)^2 }
\geq c U \, ,
\end{equation}
with a constant $c$. Since $B=\e^\sigma$, Eq.~(\ref{FhGV3})
yields,
\begin{equation}
\label{FhGV9B}
0=- \e^\sigma d\sigma + \left( F''(A) + h(\phi) G''(A) \right) dA + h'(\phi) G'(A) d\phi\, ,
\end{equation}
and we obtain,
\begin{equation}
\label{FhGV9C}
\frac{\partial A(\phi,\sigma)}{\partial \phi}
= - \frac{h'(\phi) G'(A)}{F''(A) + h(\phi) G''(A)}\, , \quad
\frac{\partial A(\phi,\sigma)}{\partial \sigma}
= - \frac{\e^\sigma}{F''(A) + h(\phi) G''(A)}\, ,
\end{equation}
which in conjunction with Eqs.~(\ref{FhGV3}) and (\ref{FhGV7}),
indicates that,
\begin{align}
\label{FhGV9D}
\frac{\partial U}{\partial \sigma} =
 - \e^{-\sigma} A - 2 \e^{-2\sigma} \left( - F(A) - h(\phi) G(A) + V (\phi) \right) \, , \quad
\frac{\partial U}{\partial \phi} = \e^{-2\sigma} V'(\phi) \, .
\end{align}
Then since $A=R$ in the original Jordan frame action, and
$B=\e^\sigma$, by using Eq.~(\ref{FhGV3}), we can rewrite the
condition (\ref{FhGV9}), as follows,
\begin{align}
\label{FhGV9F}
& \left[ \frac{1}{3}
\left\{ - \frac{R}{F'(R) + h(\phi) G'(R)}
 - \frac{ 2 \left( - F(R) - h(\phi) G(R) + V (\phi) \right)}
{ \left( F'(R) + h(\phi) G'(R) \right)^2} \right\}^2
+ \frac{\omega(\phi) \left( V' (\phi) \right)^2}{\left( F'(R) + h(\phi) G'(R) \right)^3}
\right]^{\frac{1}{2}} \nn
& \geq c \left[ \frac{R}{F'(R) + h(\phi) G'(R)}
+ \frac{ - F(R) - h(\phi) G(R) + V (\phi) }
{ \left( F'(R) + h(\phi) G'(R) \right)^2} \right] \, .
\end{align}
As an example, we consider the following model,
\begin{equation}
\label{Ex1}
F(R) = R + f_0 R^2 \, , \quad h(\phi) G(R) = h_0 \phi^{-\delta} R^\gamma \, .
\end{equation}
Here $f_0$, $h_0$, $\delta$, and $\gamma$ are constants and we may assume
$\delta>0$ and $0<\gamma<3/4$.
Then Eq.~(\ref{FhGV3}) has the following form
\begin{equation}
\label{Ex2}
B = 1 + 2 f_0 R + \gamma h_0 \phi^{-\delta} R^{\gamma - 1}\, ,
\end{equation}
and the condition~(\ref{FhGV9F}) gives
\begin{align}
\label{Ex2} & \sqrt{ \frac{1}{3} \left( R + \left( 2 - \gamma
\right) h_0 \phi^{-\delta} R^\gamma - V(\phi) \right)^2 +
\omega(\phi) \left( V'(\phi) \right)^2 \left( 1 + 2 f_0 R + \gamma
h_0 \phi^{-\delta} R^{\gamma - 1} \right) } \nn & \leq c \left(
f_0 R^2 + \left( \gamma - 1 \right) h_0 \phi^{-\delta} R^\gamma +
V(\phi) \right) \, ,
\end{align}
which constrains the value of $h_0$ and the non-minimal coupling
$h(\phi)$ in general.

\section{General Dynamics in the Jordan Frame: Inflation and Intermediate Eras\label{Sec2}}

In this section we shall discuss the dynamical evolution of the
scalar-$F(R)$ gravity gravitational system, which is governed by
the equations of motion. Also we shall show how the gravitational
equations can be used as a reconstruction technique, and in effect
an arbitrary cosmological evolution may be realized by the theory
at hand.

By varying the action~(\ref{FhGV1}) with respect to the metric, we
obtain the following equations of motion,
\begin{align}
\label{FhGV10}
0 = & \frac{1}{2}g_{\mu\nu} \left( F(R) + h(\phi) G(R) \right)
 - R_{\mu\nu} \left( F'(R) + h(\phi) G'(R) \right)
 - g_{\mu\nu} \Box \left( F'(R) + h(\phi) G'(R) \right)
+ \nabla_\mu \nabla_\nu \left( F'(R) + h(\phi) G'(R) \right) \nn
& + \frac{1}{2} \omega (\phi ) \partial_\mu \phi \partial_\nu \phi
+ \frac{1}{2} g_{\mu\nu} \left( - \frac{1}{2} \omega (\phi ) \partial_\rho \phi \partial^\rho \phi
 - V \left( \phi \right) \right)
+ \frac{\kappa^2}{2}T_{\mathrm{matter}\, \mu\nu}\, ,
\end{align}
where $T_{\mathrm{matter}\, \mu\nu}$ is the energy-momentum tensor
of the matter fluids present. On the other hand, the variation of
the action~(\ref{FhGV1}) with respect to the scalar field $\phi$
gives,
\begin{equation}
\label{FhGV11}
0 = \partial^\mu \left( \omega(\phi) \partial_\mu \phi \right)
+ h'(\phi) G(R) - V'(\phi) \, .
\end{equation}
In a spatially flat FRW universe,
\begin{equation}
\label{JGRG14}
ds^2 = - dt^2 + a(t)^2 \sum_{i=1,2,3} \left(dx^i\right)^2\, ,
\end{equation}
and by assuming that the scalar field $\phi$ depends only on the
cosmological time $t$, the $(t,t)$ and $(i,j)$ components of
(\ref{FhGV10}) have the following form,
\begin{align}
\label{FhGV12}
0 =& -\frac{1}{2}\left( F(R) + h(\phi) G(R) \right)
+ 3\left(H^2 + \dot H\right) \left( F'(R) + h(\phi) G'(R) \right)
 - 3H \frac{d}{dt} \left( F'(R) + h(\phi) G'(R) \right) \nn
& + \frac{1}{4} \omega (\phi ) {\dot\phi}^2 + \frac{1}{2} V \left( \phi \right)
+ \kappa^2 \rho \, ,\\
\label{FhGV13}
0 =& \frac{1}{2} \left( F(R) + h(\phi) G(R) \right)
 - \left(\dot H + 3H^2\right) \left( F'(R) + h(\phi) G'(R) \right)
+ \left( \frac{d^2}{dt^2} + 2 H \frac{d}{dt} \right)
\left( F'(R) + h(\phi) G'(R) \right)
 \nn
& + \frac{1}{4} \omega (\phi ) {\dot\phi}^2 - \frac{1}{2} V \left(
\phi \right) + \kappa^2 p \, ,
\end{align}
where, the Hubble rate $H$ is defined by $H=\dot a/a$ and the
scalar curvature $R$ is given by $R=12H^2 + 6\dot H$. Furthermore
$\rho$ and $p$ are the energy density and the pressure of the
matter fluids, respectively. By using the ambiguity for the
redefinition of the scalar field $\phi$, we may identify $\phi=t$.
Then Eqs.~(\ref{FhGV12}) and (\ref{FhGV13}) can be rewritten as,
\begin{align}
\label{FhGV14}
\omega (\phi ) =& - 2 \left. \left\{ 2 \dot H \left( F'(R) + h(\phi) G'(R) \right)
+ \left( \frac{d^2}{dt^2} - H \frac{d}{dt} \right)
\left( F'(R) + h(\phi) G'(R) \right)
+ \kappa^2 \left( \rho + p \right) \right\} \right|_{t=\phi} \, ,\\
\label{FhGV15}
V \left( \phi \right) =& \Bigl[ F(R) + h(\phi) G(R)
 - \left(4 \dot H + 6 H^2\right) \left( F'(R) + h(\phi) G'(R) \right)
+ \left( \frac{d^2}{dt^2} + 5 H \frac{d}{dt} \right)
\left( F'(R) + h(\phi) G'(R) \right) \nn
& \left. - \kappa^2 \left( \rho - p \right) \Bigr] \right|_{t=\phi}\, ,
\end{align}
Then given an arbitrary cosmological evolution $H(t)$, by using
the equation of state and the continuity equation, we may find the
explicit $t$ dependencies of $\rho $ and $p $. Then the right hand
side of both Eqs.~(\ref{FhGV14}) and (\ref{FhGV15}) is expressed
as a function of the cosmic time $t$. Since $\phi=t$, by replacing
$t$ with $\phi$ in the right hand side of both Eqs.~(\ref{FhGV14})
and (\ref{FhGV15}), we obtain the explicit forms of $\omega(\phi)$
and $V(\phi)$ as functions of the scalar field $\phi$. This
indicates that an arbitrary expansion history of the Universe
described by $H=H(t)$ can be realized by choosing $\omega(\phi)$
and $V(\phi)$ to satisfy Eqs.~(\ref{FhGV14}) and (\ref{FhGV15})
for any form of $F(R)$, $G(R)$, and $h(\phi)$. As an example, we
consider the model in which the scale factor is,
\begin{equation}
\label{FhGVex1} a = \mathcal{A} (t) \equiv a_0 \e^{ \frac{\left(
H_I - H_L \right) t}{1 + H_T t} + H_L t }\, ,
\end{equation}
where $a_0$, $H_I$, $H_T$, and $H_L$ are dimensionful constants.
Then, Eq.~(\ref{FhGVex1}) yields,
\begin{equation}
\label{FhGVex2}
H = \mathcal{H}(t) \equiv \frac{\left( H_I - H_L \right) }{ \left(1 + H_T t\right)^2 } + H_L \, .
\end{equation}
In effect, cosmic times for which $t\ll \frac{1}{H_T}$, $H\sim
H_I$, correspond to the inflationary era in the early Universe and
cosmic times for which $t\gg \frac{1}{H_T}$, $H\sim H_L$,
correspond to the accelerated expansion of the late Universe. We
also find that the scalar curvature $R$ is given by,
\begin{equation}
\label{FhGVex3}
R = \mathcal{R} (t)
\equiv 12 \left\{ \left( \frac{\left( H_I - H_L \right) }{ \left(1 + H_T t\right)^2 }
+ H_L \right)^2 - \frac{\left( H_I - H_L \right) H_T }{ \left(1 + H_T t\right)^3 }
\right\} \, .
\end{equation}
When the matter has a constant EoS parameter $w\equiv
\frac{p}{\rho}$, $\rho$ is given by $\rho = \rho_0 a^{-3 \left(
1+w \right)}$ with a constant $\rho_0$. Then Eqs.~(\ref{FhGV14})
and (\ref{FhGV15}) indicate that the Universe whose evolution in
the expansion is given by (\ref{FhGVex1}) is realized by choosing,
\begin{align}
\label{FhGV14ex}
\omega (\phi ) =& - 4 \mathcal{H}' \left(\phi \right)
\left( F'\left(\mathcal{R} \left(\phi \right) \right)
+ h(\phi) G'\left(\mathcal{R}\left(\phi\right) \right) \right)
 - 2\left( \frac{d^2}{d\phi^2} - \mathcal{H} \left( \phi \right) \frac{d}{d\phi} \right)
\left( F'\left( \mathcal{R} \left( \phi \right) \right)
+ h(\phi) G' \left( \mathcal{R} \left( \phi \right) \right) \right) \nn
& -2 \kappa^2 \left( 1 + w \right) \rho_0
\left( \mathcal{A} \left(\phi \right) \right)^{-3(1+w)} \, ,\\
\label{FhGV15ex}
V \left( \phi \right) =& F \left( \mathcal{R} \left( \phi \right) \right)
+ h(\phi) G \left( \mathcal{R} \left( \phi \right) \right)
 - \left(4 \mathcal{H}' \left( \phi \right) + 6 \mathcal{H} \left(\phi\right)^2\right)
\left( F' \left( \mathcal{R} \left( \phi \right) \right)
+ h(\phi) G' \left( \mathcal{R} \left( \phi \right) \right) \right) \nn
& + \left( \frac{d^2}{d\phi^2} + 5 \mathcal{H} \left( \phi \right) \frac{d}{d\phi} \right)
\left( F' \left( \mathcal{R} \left( \phi \right) \right)
+ h(\phi) G' \left( \mathcal{R} \left( \phi \right) \right) \right)
 - \kappa^2 \left( 1 - w \right) \rho_0 \left( \mathcal{A} \left(\phi \right)
\right)^{-3(1+w)}  \, .
\end{align}
Here the functions $F(R)$, $G(R)$, and $h(\phi)$ can be arbitrary but we may choose
them as in (\ref{Ex1}).

\section{Gravitational Waves of the Axion-$F(R)$ Gravity Theory\label{Sec3}}

We now consider the gravitational wave based on the action~(\ref{FhGV7})
in the Einstein frame by considering the
perturbation of the background ${\tilde g}_{\mu\nu} = {\tilde
g^{(0)}}_{\mu\nu}$ as ${\tilde g}_{\mu\nu} = {\tilde
g^{(0)}}_{\mu\nu} + {\tilde h}_{\mu\nu}$ in the Einstein equation,
\begin{equation}
\label{FhGV16}
{\tilde R}_{\mu\nu} - \frac{1}{2} {\tilde g}_{\mu\nu} {\tilde R}
= 3 \partial_\mu \sigma \partial_\nu \sigma
+ \e^{-\sigma} \omega (\phi ) \partial_\mu \phi \partial_\nu \phi
+ {\tilde g}_{\mu\nu} \left( - \frac{3}{2} \partial_\rho \sigma \partial^\rho \sigma
 - \frac{1}{2} \e^{-\sigma} \omega (\phi ) \partial_\rho \phi \partial^\rho \phi
 - U \left( \phi, \sigma \right) \right)
+ \kappa^2 {\tilde T}_{\mathrm{matter}\, \mu\nu} \, .
\end{equation}
Here the matter energy momentum tensor ${\tilde
T}_{\mathrm{matter}}^{\mu\nu}$ in the Einstein frame is,
\begin{equation}
\label{FhGV17O}
{\tilde T}_{\mathrm{matter}}^{\mu\nu} \equiv \frac{2}{\sqrt{- \tilde g}}
\frac{\partial \left( \sqrt{- \tilde g} \e^{-2\sigma} \mathcal{L}_\mathrm{matter}
\left( \e^{-\sigma} {\tilde g}_{\mu\nu}, \Phi_i \right) \right)}
{\partial {\tilde g}_{\mu\nu}} \, .
\end{equation}
If the matter fluids couple minimally with gravity, that is, if
the matter Lagrangian $ \mathcal{L}_\mathrm{matter} \left(
\e^{-\sigma} {\tilde g}_{\mu\nu}, \Phi_i \right) =
\mathcal{L}_\mathrm{matter} \left( g_{\mu\nu}, \Phi_i \right)$
does not include any derivative of the metric $g_{\mu\nu}$, the
matter energy momentum tensor ${\tilde
T}_{\mathrm{matter}}^{\mu\nu}$ in the Einstein frame is related
with the matter energy momentum tensor
$T_{\mathrm{matter}}^{\mu\nu}$ in the original Jordan frame as
${\tilde T}_{\mathrm{matter}}^{\mu\nu} = \e^{-3\sigma}
T_{\mathrm{matter}}^{\mu\nu}$, that is, ${\tilde
T}_{\mathrm{matter}\, \mu\nu} = \e^{-\sigma} T_{\mathrm{matter}\,
\mu\nu}$. When we consider the gravitational wave, we often use
the transverse and traceless gauge conditions. Since we are
considering the scale transformation~(\ref{FhGV6}), if
$h_{\mu\nu}$, which is defined by the fluctuation from the
background metric $g_{\mu\nu} = g^{(0)}_{\mu\nu}$ as $g_{\mu\nu} =
g^{(0)}_{\mu\nu} + h_{\mu\nu}$ in the original frame in
the action~(\ref{FhGV1}), satisfies the transverse and traceless gauge
conditions,
\begin{equation}
\label{FhGV17} \nabla^\mu h_{\mu\nu} = g^{(0)\, \mu\nu} h_{\mu\nu}
= 0 \, .
\end{equation}
However, the scale transformed fluctuation ${\tilde h}_{\mu\nu} =
\e^\sigma h_{\mu\nu}$ does not always satisfies the first
condition in (\ref{FhGV17}), although the second condition is
trivially satisfied, ${\tilde g}^{(0)\, \mu\nu} {\tilde
h}_{\mu\nu} = \e^{-\sigma} g^{(0)\, \mu\nu} \e^\sigma h_{\mu\nu} =
g^{(0)\, \mu\nu} h_{\mu\nu} = 0$. For the first condition in
(\ref{FhGV17}), under the scale transformation, we find,
\begin{align}
\label{FhGV18}
& \tilde{\nabla}^\mu \tilde{h}_{\mu}^{\ \nu}
= \e^{-\sigma} \nabla^\mu h_{\mu\nu}
+ 4 \e^{-\sigma} g^{(0)\, \mu\tau} g^{(0)\, \nu\rho} \sigma_{,\tau} h_{\mu\rho}
 - \e^{-\sigma} g^{(0)\, \nu\rho}\sigma_{,\rho} g^{(0)\, \mu\tau} h_{\mu\tau}
= 4 \e^{-\sigma} g^{(0)\, \mu\tau} g^{(0)\, \nu\rho} \sigma_{,\tau} h_{\mu\rho} \, .
\end{align}
Then if we assume that a homogeneous and isometric background
metric, and therefore $\sigma$ only depends on the cosmological
time $t$ and also $g^{(0)}_{ti}=0$, then if we consider the
perturbation with $h_{t\mu}=0$ since we are considering the
massless spin 2 mode, we find,
\begin{equation}
\label{FhGV19} \tilde{\nabla}^\mu \tilde{h}_{\mu}^{\ \nu} =
{\tilde g}^{(0)\, \mu\nu} {\tilde h}_{\mu\nu} = 0\, .
\end{equation}
Therefore the gauge conditions in (\ref{FhGV17}) for the graviton are not changed
by the scale transformation,
\begin{equation}
\label{FhGV21}
{\tilde\nabla}^\mu {\tilde h}_{\mu\nu} = {\tilde g}^{(0)\, \mu\nu} {\tilde h}_{\mu\nu} = 0 \, .
\end{equation}
Then under the condition~(\ref{FhGV21}), the equation for the
gravitational wave can be written as follows,
\begin{align}
\label{FhGV22}
0 =& \frac{1}{2\kappa^2}\left(- \frac{1}{2}\left( - {\tilde \Box}^{(0)} {\tilde h}_{\mu\nu}
 - 2{\tilde R}^{(0)\, \lambda\ \rho}_{\ \ \ \ \ \nu\ \mu} {\tilde h}_{\lambda\rho}
+ {\tilde R}^{(0)\, \rho}_{\ \ \ \ \ \mu}{\tilde h}_{\rho\nu}
+ {\tilde R}^{(0)\, \rho}_{\ \ \ \ \ \nu}{\tilde h}_{\rho\mu} \right)
+ \frac{1}{2} R^{(0)} h_{\mu\nu}
 - \frac{1}{2}{\tilde g}^{(0)}_{\mu\nu} {\tilde h}_{\rho\sigma}
{\tilde R}^{(0)\, \rho\sigma} \right) \nn
& + {\tilde h}_{\mu\nu} \left( - \frac{3}{2} \partial_\rho \sigma \partial^\rho \sigma
 - \frac{1}{2} \e^{-\sigma} \omega (\phi ) \partial_\rho \phi \partial^\rho \phi
 - U \left( \phi, \sigma \right) \right)
- {\tilde g}^{(0)}_{\mu\nu} \left( \frac{3}{2} \partial^\rho \sigma \partial^\tau \sigma
 - \frac{1}{2} \e^{-\sigma} \omega (\phi ) \partial^\rho \phi \partial^\tau \phi
\right) {\tilde h}_{\rho\tau} \nn
& + \frac{1}{2} \frac{\partial {\tilde T}_{\mathrm{matter}\, \mu\nu}}
{\partial {\tilde g}_{\rho\tau}}
{\tilde h}_{\rho\tau} \, .
\end{align}
We are now interested in the massless spin two mode, which
satisfies,
\begin{equation}
\label{FhGV23}
{\tilde h}_{it}={\tilde h}_{ti}=h_{it}=h_{ti}=0\, , \quad
\sum_{i=1,2,3} {\tilde h}_{ii}=\sum_{i=1,2,3} h_{ii}=0 \, , \quad i=1,2,3\, ,
\quad {\tilde h}_{tt}=h_{tt}=0 \, .
\end{equation}
In the spatially flat FRW universe in the Einstein frame,
\begin{equation}
\label{JGRG14E}
d{\tilde s}^2 \equiv \e^{\sigma} ds^2 = - d{\tilde t}^2
+ {\tilde a} \left( \tilde t \right)^2 \sum_{i=1,2,3} \left(dx^i\right)^2\, ,
\end{equation}
where $d{\tilde t}\equiv \e^{\frac{\sigma}{2}} dt$ and $\tilde a
\left( \tilde t \right) \equiv \e^{\frac{\sigma}{2}} a(t)$, due to
the isometry in the spacial part, we may assume,
\begin{equation}
\label{FhGV24}
\frac{\partial {\tilde T}_{\mathrm{matter}\, tt}}{\partial{\tilde g}_{ij}} \propto \delta^{ij} \, , \quad
\frac{\partial {\tilde T}_{\mathrm{matter}\, tk}}{\partial {\tilde g}_{ij}}
= \frac{\partial {\tilde T}_{\mathrm{matter}\, kt}}{\partial {\tilde g}_{ij}} = 0 \, .
\end{equation}
We may further assume that the matter energy-momentum tensor~(\ref{FhGV17O})
in the Einstein frame has the following form as in the perfect fluid,
\begin{equation}
\label{FhGV25}
{\tilde T}_{\mathrm{matter}\, \mu\nu} =\tilde\rho {\tilde U}_\mu {\tilde U}_\nu
+ \tilde p {\tilde\gamma}_{\mu\nu} \, .
\end{equation}
Here $\left( \tilde U^\mu \right)$ is the four velocity of the matter
fluid and we now assume $\tilde U^0=1$ and $\tilde U^i=0$. In
Eq.~(\ref{FhGV25}), ${\tilde\gamma}_{\mu\nu}$ is the projection
tensor to the spatial directions perpendicular to $\tilde U^\mu$,
\begin{equation}
\label{FhGV26}
{\tilde\gamma}_{\mu\nu}={\tilde g}_{\mu\nu} + {\tilde U}_\mu {\tilde U}_\nu\, .
\end{equation}
We now also assume that the matter fluid minimally couples with
the metric ${\tilde g}_{\mu\nu}$, that is, the coupling between
the matter fluids and the metric does not include the derivative
of the metric. Then, under the perturbation ${\tilde g}_{\mu\nu} =
{\tilde g}^{(0)}_{\mu\nu} + {\tilde h}_{\mu\nu}$, we find,
\begin{equation}
\label{FhGV27}
\delta \tilde \rho = {\tilde \rho}^{\mu\nu} {\tilde h}_{\mu\nu}\, , \quad
\delta \tilde p = {\tilde p}^{\mu\nu} {\tilde h}_{\mu\nu}\, .
\end{equation}
On the other hand, the variation of ${\tilde U}_\mu$ is given by
using the condition ${\tilde U}^\mu {\tilde U}_\mu = -1$, that is,
\begin{equation}
\label{FhGV29}
0 = 2 \left( \delta {\tilde U}^\mu \right)
+ {\tilde U}^\mu {\tilde U}^\nu {\tilde h}_{\mu\nu}
= {\tilde U}^\mu \left( 2 {\tilde g}^{(0)}_{\mu\nu} \delta {\tilde U}^\nu
+ {\tilde h}_{\mu\nu} {\tilde U}^\nu \right)\, ,
\end{equation}
and hence we have,
\begin{equation}
\label{FhGV30}
0 = 2 \left( \delta {\tilde U}^\mu \right)
+ {\tilde U}^\mu {\tilde U}^\nu {\tilde h}_{\mu\nu}
= {\tilde U}^\mu \left( 2 {\tilde g}^{(0)}_{\mu\nu} \delta {\tilde U}^\nu
+ {\tilde h}_{\mu\nu} {\tilde U}^\nu \right)\, ,
\end{equation}
or equivalently,
\begin{equation}
\label{FhGV31} \delta {\tilde U}^\mu = - \frac{1}{2} {\tilde
g}^{(0)\, \mu\rho} \left( {\tilde h}_{\rho\nu} {\tilde U}^\nu +
l_\rho \right)\, .
\end{equation}
Note that the arbitrary vector $l_\mu$ satisfies ${\tilde U}^\mu
l_\mu = 0$, but we choose $l_\mu=0$ by assuming the isometry. Due
to the isometry, we may assume $\rho^{ij}$ and $p^{ij}$ are
proportional to $\delta^{ij}$, Then the $(t,t)$ and $(t,i)$
components of equation for gravitational wave (\ref{FhGV22}) are
trivially satisfied and the $(i,j)$ component is given by,
\begin{align}
\label{FhGV32}
0 =& \frac{1}{2\kappa^2}\left( \frac{1}{2}
\left( - \partial_{\tilde t}^2 {\tilde h}_{ij} + {\tilde a}^{-2} \triangle {\tilde h}_{ij} \right)
+ \left( 3 \frac{d \tilde H}{d\tilde t} + 4 {\tilde H}^2 \right) {\tilde h}_{ij}
+ {\tilde h}_{ij} \left( \frac{3}{2} \left( \frac{d \sigma}{d\tilde t} \right)^2
+ \frac{1}{2} \e^{-\sigma} \omega (\phi )
\left( \frac{d \phi}{d\tilde t} \right)^2
 - U \left( \phi, \sigma \right) \right) \right)
+ \frac{1}{2} \tilde p {\tilde h}_{ij} \nn
=& \frac{1}{2\kappa^2}\left( \frac{1}{2}
\left( - \partial_{\tilde t}^2 {\tilde h}_{ij} + {\tilde a}^{-2} \triangle {\tilde h}_{ij} \right)
+ \left( 3 \frac{d \tilde H}{d\tilde t} + 4 {\tilde H}^2 \right) {\tilde h}_{ij}
+ {\tilde h}_{ij} \left( \frac{3}{2} \left( \frac{d \sigma}{d\tilde t} \right)^2
+ \frac{1}{2} \e^{-2\sigma} \omega (\phi )
 - U \left( \phi, \sigma \right) \right) \right)
+ \frac{1}{2} \tilde p {\tilde h}_{ij} \, .
\end{align}
Here we have used $\left( \frac{d \phi}{d\tilde t} \right)^2
=\left( \frac{dt}{d\tilde t} \right)^2 = \e^{-\sigma}$. We may
rewrite Eq.~(\ref{FhGV32}) in the original Jordan frame. Since,
${\tilde h}_{\mu\nu} = \e^\sigma h_{\mu\nu}$ and $\tilde a =
\e^{\frac{\sigma}{2}} a$, we find,
\begin{align}
\label{FhGV33}
\frac{\partial}{\partial \tilde t} =& \e^{-\frac{\sigma}{2}} \frac{\partial}{\partial t} \, , \quad
\frac{\partial^2}{\partial \tilde t^2} = \e^{-\sigma}
\left( \frac{\partial^2}{\partial t^2 } - \frac{1}{2}\frac{d\sigma}{dt} \frac{\partial}{\partial t}
\right) \, , \nn
\tilde H =& \frac{1}{\tilde a} \frac{d \tilde a}{d\tilde t}
= \e^{-\frac{\sigma}{2}} \left( \frac{1}{2} \frac{d \sigma }{d t} + H \right) \, , \quad
\frac{d \tilde H}{d\tilde t} = \e^{-\tilde \sigma} \left( \frac{1}{2} \frac{d^2 \sigma}{dt^2}
+ \frac{1}{4} \left( \frac{d \sigma}{dt} \right)^2 + \frac{d H}{dt} \right) \, , \nn
\frac{\partial^2 {\tilde h}_{ij}}{\partial \tilde t^2} =&
\frac{\partial^2 h_{ij} }{\partial t^2} + \frac{d^2 \sigma}{dt^2} h_{ij}
+ \left( \frac{d\sigma}{dt} \right)^2 h_{ij}
+ 2 \frac{d \sigma}{dt} \frac{\partial h_{ij}}{\partial t}
 - \frac{1}{2} \left( \frac{d \sigma}{dt} \right)^2 \frac{\partial h_{ij}}{\partial t}
 - \frac{1}{2}\frac{d\sigma}{dt} \frac{\partial h_{ij}}{\partial t} \nn
=& \frac{\partial^2 h_{ij} }{\partial t^2} + \frac{d^2 \sigma}{dt^2} h_{ij}
+ \frac{1}{2} \left( \frac{d\sigma}{dt} \right)^2 h_{ij}
+ \frac{3}{2} \frac{d \sigma}{dt} \frac{\partial h_{ij}}{\partial t} \, ,
\end{align}
Then, due to the fact that $\tilde p = \e^{-2\sigma} p$ from
(\ref{FhGV25}) (${\tilde T}_{\mathrm{matter}\, \mu\nu} =
\e^{-\sigma} T_{\mathrm{matter}\, \mu\nu}$ and
${\tilde\gamma}_{\mu\nu} = \e^\sigma \gamma_{\mu\nu}$),
Eq.~(\ref{FhGV32}) is rewritten as follows,
\begin{align}
\label{FhGV34}
0 =& \frac{1}{2\kappa^2}\left( \frac{1}{2}
\left( - \partial_t^2 h_{ij} + \frac{3}{2} \dot\sigma \partial_t h_{ij}
+ \left( \ddot\sigma + {\dot\sigma}^2 \right) h_{ij} + a^{-2} \triangle h_{ij} \right) \right. \nn
& \left. + \left( 3 \dot H + 4 H^2 + \frac{3}{2} \ddot \sigma + \frac{13}{4} {\dot \sigma}^2
+ 4 \dot\sigma H + \frac{1}{2} \omega (\phi ) {\dot\phi}^2
 - \e^{\sigma} U \left( \phi, \sigma \right) \right) h_{ij} \right)
+ \frac{1}{2}\e^{-\sigma} p h_{ij} \, .
\end{align}
We should note that $\e^\sigma$ is given in Eq.~(\ref{FhGV3}) and
$U \left( \phi, \sigma \right)$ is given in Eq.~(\ref{FhGV7}).
Then in terms of the Jordan frame, we find,
\begin{align}
\label{FhGV35}
\sigma = & \ln \left( F'(R) + h(\phi) G'(R) \right) \, , \nn
U \left( \phi, \sigma \right) = & \e^{-\sigma} R
+ \e^{-2\sigma} \left\{ - F\left( R \right) - h(\phi) G \left( R \right)
+ V (\phi) \right\} \nn
= & \frac{R \left( F'(R) + h(\phi) G'(R) \right)
 - F\left( R \right) - h(\phi) G \left( R \right)
+ V (\phi) }{\left( F'(R) + h(\phi) G'(R) \right)^2} \, .
\end{align}
By choosing $\phi=t$ and using (\ref{FhGV14}) and (\ref{FhGV15}),
we obtain,
\begin{align}
\label{FhGV36}
\frac{1}{2} & \omega (\phi ) {\dot\phi}^2 - \e^{\sigma} U \left( \phi, \sigma \right) \nn
=& - \left( F'(R) + h(\phi) G'(R) \right)^{-1}
\left( 2 \dot H \left( F'(R) + h(\phi) G'(R) \right)
+ \left( \frac{d^2}{dt^2} - H \frac{d}{dt} \right)
\left( F'(R) + h(\phi) G'(R) \right)
+ \kappa^2 \left( \rho + p \right) \right. \nn
& - \left( \left( 12 H^2 + 6 \dot H \right) \left( F'(R) + h(\phi) G'(R) \right)
 - \left(4 \dot H + 6 H^2\right) \left( F'(R) + h(\phi) G'(R) \right) \right. \nn
& \left. \left. + \left( \frac{d^2}{dt^2} + 5 H \frac{d}{dt} \right)
\left( F'(R) + h(\phi) G'(R) \right) - \kappa^2 \left( \rho - p
\right) \right) \right) \nn
=& - \left( F'(R) + h(\phi) G'(R) \right)^{-1}
\left( - 6 \left( H^2 + H \frac{d}{dt} \right) \left( F'(R) + h(\phi) G'(R) \right)
+ 2 \kappa^2 \rho \right) \nn
=& 6 H^2 + 6 H \dot\sigma + 2 \kappa^2 \e^{-\sigma} \rho \, ,
\end{align}
and we can further rewrite Eq.~(\ref{FhGV34}) as follows,
\begin{equation}
\label{FhGV37}
0 = \frac{1}{2\kappa^2}\left( \frac{1}{2}
\left( - \partial_t^2 h_{ij} + \frac{3}{2} \dot\sigma \partial_t h_{ij}
+ a^{-2} \triangle h_{ij} \right)
+ \left( 3 \dot H + 10 H^2 + 2 \ddot \sigma + \frac{17}{4} {\dot \sigma}^2
+ 10 \dot\sigma H \right) h_{ij} \right)
+ \frac{1}{2}\e^{-\sigma} \left( \rho + p \right) h_{ij} \, .
\end{equation}
As an example, we consider the quasi-de Sitter spacetime case,
\begin{equation}
\label{FhGV37}
H=H_0 + H_1 \left( t - t_0 \right) 
 \, .
\end{equation}
Here $H_0$, $H_1$, and $t_0$ are constants and we assume the
second term in (\ref{FhGV37}) is much smaller than the first one,
and the third term is also much smaller than the second term, $H_0
\gg \left| H_1 \left( t - t_0 \right) \right| \gg \left| H_2
\left( t - t_0 \right)^2 \right|$ by assuming $H_0$ is positive.
We may also assume that $H_0^2 \sim H_1$. 
Then we find,
\begin{equation}
\label{FhGV38}
R = 12 H_0^2 + 6 H_1 + 
24 H_0 H_1 
\left( t - t_0 \right) + \mathcal{O} \left( \left( t - t_0 \right)^2 \right) \, .
\end{equation}
By further assuming $\left| t - t_0 \right| \ll \left| t_0
\right|$, we find,
\begin{align}
\label{FhGV38}
\sigma =&
\sigma_0 + \sigma_1 \left( t - t_0 \right)
+ \mathcal{O} \left( \left( t - t_0 \right)^2 \right) \nn
\equiv& \ln \left( F'\left( R_0 \right)
+ h \left( t_0 \right) G' \left( R_0 \right) \right)
+ \frac{ \left( 24 H_0 H_1 F''\left( R_0 \right)
+ h \left( t_0 \right) G'' \left( R_0 \right) \right) + h' \left( t_0 \right)  G' \left( R_0 \right) }
{ F'\left( R_0 \right) + h \left( t_0 \right) G' \left( R_0 \right)} \left( t - t_0 \right) \nn
& + \mathcal{O} \left( \left( t - t_0 \right)^2 \right) \, .
\end{align}
Then Eq.~(\ref{FhGV37}) can be approximated as follows,
\begin{equation}
\label{FhGV39}
0 = \frac{1}{2\kappa^2}\left( \frac{1}{2}
\left( - \partial_t^2 h_{ij} + \frac{3}{2} \sigma_1 \partial_t h_{ij}
+ \triangle h_{ij} \right)
+ \left( 3 H_1 + 10 H_0^2 + \frac{17}{4} \sigma_1^2
+ 10 \sigma_1 H_0 \right) h_{ij} \right)
+ \frac{1}{2}\e^{-\sigma_0} \left( \rho + p \right) h_{ij} \, .
\end{equation}
Here we have neglected $\ddot\sigma$ and we put $a=1$ because we
now considering the propagation of the gravitational wave for a
time scale much shorter than the scale of the expansion of the
Universe. For simplicity, we neglect the contribution from the
matter fluids by putting $\rho=p=0$. Then we may separate the
variables by assuming $h_{ij} = \e^{i\bm{k}\cdot\bm{x}} \hat
h_{ij} (t)$. Then from Eq.~(\ref{FhGV39} we get,
\begin{equation}
\label{FhGV40} 0 = \frac{1}{2\kappa^2}\left( \frac{1}{2} \left( -
\partial_t^2 h_{ij} + \frac{3}{2} \sigma_1 \partial_t h_{ij}
\right) + \left(- \frac{k^2}{2} + 3 H_1 + 10 H_0^2 + \frac{17}{4}
\sigma_1^2 + 10 \sigma_1 H_0 \right) h_{ij} \right) +
\frac{1}{2}\e^{-\sigma_0} \left( \rho + p \right) h_{ij} \, ,
\end{equation}
where $k^2 \equiv \bm{k}\cdot\bm{k}$. By assuming $\hat h_{ij} (t)
\propto \e^{-i\omega t}$, we find,
\begin{equation}
\label{FhGV40}
\omega = i \frac{3}{4} \sigma_1 \pm \sqrt{ - \frac{9}{16}\sigma_1^2
+ k^2 - 6 H_1 - 20 H_0^2 - \frac{17}{2} \sigma_1^2
 - 20 \sigma_1 H_0 } \, .
\end{equation}
For large wavenumbers, $k^2 \gg \left| \frac{9}{16}\sigma_1^2 + 6
H_1 + 20 H_0^2 + \frac{17}{2} \sigma_1^2  20 \sigma_1 H_0
\right|$, we find $\omega\sim k$, and therefore there is no change
in the propagation speed of the gravitational wave as in the
standard $F(R)$ gravity case \cite{Nojiri:2017hai}. If $\sigma_1 >
0$, however, the gravitational wave is enhanced and if $\sigma_1
<0$, dissipation of the gravity wave occurs. The situation is very
similar to the propagation in a viscous fluid
\cite{Brevik:2019yma}. The enhancement or the dissipation of the
gravity wave occurs due to the term $\frac{3}{2} \sigma_1
\partial_t h_{ij}$ in Eq.~(\ref{FhGV40}), which includes the first
derivative of $h_{ij}$. In Eq.~(\ref{FhGV32}) in the Einstein
frame, such a first derivative term does not appear. Then the
enhancement or the dissipation of the gravity wave occurs as an
effect originating from the scale transformation, ${\tilde
h}_{\mu\nu} = \e^\sigma h_{\mu\nu}$. The effect of the enhancement
or the dissipation from the scale transformation has been also
found in the standard $F(R)$ gravity \cite{Capozziello:2017vdi}.
For the recently observed gravitational waves,
\cite{Abbott:2016blz,Abbott:2016nmj,Abbott:2017vtc,Abbott:2017oio,
Abbott:2017gyy,TheLIGOScientific:2017qsa}, the distances between
the sources and the earth are about a few hundreds Mpc. Because no
dissipation or enhancement of the gravitational waves has been
observed we find the following constraint in the present Universe,
\begin{equation}
\label{FhGV41} \left| \sigma_1 \right| \ll \left( 10^3\,
\mathrm{Mpc}\right)^{-1}\, .
\end{equation}

\section{Energy Momentum Tensor, Energy Conditions and Constraints of the Axion-like Particle Minimal Coupling $h(\phi)$\label{Sec4}}

We now consider the conserved quantities like energy-momentum
tensors as in \cite{Capozziello:2018wul}, and also we shall
constrain the parameter $\delta$ appearing in the non-minimal
coupling of $h(\phi)$ to the scalar curvature term $G(R)$. In
(\ref{FhGV10}), the energy-momentum tensor $T_{\mathrm{matter}\,
\mu\nu}$ of the matter fluids is, of course, conserved,
$\nabla^\mu T_{\mathrm{matter}\, \mu\nu}=0$. As in the standard
$F(R)$ gravity, the Bianchi identity indicates that the following
quantity is conserved,
\begin{equation}
\label{FhGV42}
T_{F(R)\, \mu\nu} \equiv \frac{2}{\kappa^2} \left( \frac{1}{2}g_{\mu\nu} F(R)
 - R_{\mu\nu} F'(R)  - g_{\mu\nu} \Box F'(R)
+ \nabla_\mu \nabla_\nu F'(R) \right) \, .
\end{equation}
Therefore by using (\ref{FhGV10}), we may define the conserved
energy momentum tensor for the scalar field $\phi$ as follows,
\begin{align}
\label{FhGV43}
T_{\phi\, \mu\nu} \equiv& \frac{2}{\kappa^2} \left(
\frac{1}{2}g_{\mu\nu} \left( h(\phi) G(R) \right)  - R_{\mu\nu} h(\phi) G'(R)
 - g_{\mu\nu} \Box \left( h(\phi) G'(R) \right)
+ \nabla_\mu \nabla_\nu \left( h(\phi) G'(R) \right) \right. \nn
& \left. + \frac{1}{2} \omega (\phi ) \partial_\mu \phi \partial_\nu \phi
+ \frac{1}{2} g_{\mu\nu} \left( - \frac{1}{2} \omega (\phi )
\partial_\rho \phi \partial^\rho \phi - V \left( \phi \right) \right) \right) \, .
\end{align}
Furthermore Eq.~(\ref{FhGV10}) can be rewritten as follows,
\begin{align}
\label{FhGV44}
R_{\mu\nu} - \frac{1}{2} g_{\mu\nu} R
= \frac{1}{F'(R) + h(\phi) G'(R)} & \left\{
\frac{1}{2}g_{\mu\nu} \left( F(R) + h(\phi) G(R)
 - F'(R) - h(\phi) G'(R) \right) \right. \nn
& - g_{\mu\nu} \Box \left( F'(R) + h(\phi) G'(R) \right)
+ \nabla_\mu \nabla_\nu \left( F'(R) + h(\phi) G'(R) \right) \nn
& \left. + \frac{1}{2} \omega (\phi ) \partial_\mu \phi \partial_\nu \phi
+ \frac{1}{2} g_{\mu\nu} \left( - \frac{1}{2} \omega (\phi ) \partial_\rho \phi \partial^\rho \phi
 - V \left( \phi \right) \right)
+ \frac{\kappa^2}{2}T_{\mathrm{matter}\, \mu\nu} \right\} \, .
\end{align}
Then by using the Bianchi identity, we find the following conserved tensor,
\begin{align}
\label{FhGV45}
{\hat T}_{\mu\nu}
= \frac{2}{\kappa^2 \left( F'(R) + h(\phi) G'(R) \right)} & \left\{
\frac{1}{2}g_{\mu\nu} \left( F(R) + h(\phi) G(R)
 - F'(R) - h(\phi) G'(R) \right) \right. \nn
& - g_{\mu\nu} \Box \left( F'(R) + h(\phi) G'(R) \right)
+ \nabla_\mu \nabla_\nu \left( F'(R) + h(\phi) G'(R) \right) \nn
& \left. + \frac{1}{2} \omega (\phi ) \partial_\mu \phi \partial_\nu \phi
+ \frac{1}{2} g_{\mu\nu} \left( - \frac{1}{2} \omega (\phi )
\partial_\rho \phi \partial^\rho \phi
 - V \left( \phi \right) \right)
+ \frac{\kappa^2}{2}T_{\mathrm{matter}\, \mu\nu} \right\} \, .
\end{align}
We should note that any linear combination of
$T_{\mathrm{matter}\, \mu\nu}$, $T_{F(R)\, \mu\nu}$, $T_{\phi\,
\mu\nu}$, and ${\hat T}_{\mu\nu}$, that is, ${\bar T}_{\mu\nu}
\equiv c_1 T_{\mathrm{matter}\, \mu\nu} + c_2 T_{F(R)\, \mu\nu} +
c_3 T_{\phi\, \mu\nu} + c_4 \tilde T_{\mu\nu}$ with constants
$c_1$, $c_2$, $c_3$, and $c_4$, is conserved. On the other hand,
in the Einstein frame, the energy-momentum tensor ${\tilde
T}_{\mathrm{matter}\, \mu\nu}$ is not conserved but
Eq.~(\ref{FhGV16}) indicates that the following quantity is
conserved,
\begin{equation}
\label{FhGV46}
\mathcal{T}_{\mu\nu}
= \frac{1}{\kappa^2} \left\{ 3 \partial_\mu \sigma \partial_\nu \sigma
+ \e^{-\sigma} \omega (\phi ) \partial_\mu \phi \partial_\nu \phi
+ {\tilde g}_{\mu\nu} \left( - \frac{3}{2} \partial_\rho \sigma \partial^\rho \sigma
 - \frac{1}{2} \e^{-\sigma} \omega (\phi ) \partial_\rho \phi \partial^\rho \phi
 - U \left( \phi, \sigma \right) \right) \right\}
+ {\tilde T}_{\mathrm{matter}\, \mu\nu} \, .
\end{equation}
In the standard $F(R)$ gravity, when $F(R)$ behaves as $F(R)
\propto R^m$, if we include the contributions from the matter with
a constant EoS parameter $w$, for a flat FRW universe~(\ref{JGRG14}),
the solution is given by,
\begin{equation}
\label{M8}
a \propto t^{h_0} \, ,\quad h_0\equiv \frac{2m}{3(1+w)} \, .
\end{equation}
Then the effective EoS parameter is,
\begin{equation}
\label{JGRG20} w_\mathrm{eff} \equiv -1 - \frac{2\dot
H}{3H^2}=-1+\frac{2}{3h_0} = -1 + \frac{w+1}{m}\, .
\end{equation}
Then even if $w>-1$, when $m<0$, all the energy conditions are
effectively not satisfied for the energy momentum tensor
correspond to (\ref{FhGV45}). However, in the axion-$F(R)$ gravity
model, all the curvature related terms, namely $F(R)$ and $G(R)$,
contain positive powers of the curvature, hence the energy
conditions are satisfied. Let us elaborate on this in a more
detailed manner. When the curvature related terms dominate the
evolution, namely at early and late-times, the axion has two
different behaviors. Particularly, for early times, the axion is
frozen, and its average EoS parameter is $w=-1$, recalling that
the axion is the sole matter content of the
model~(\ref{mainaction}), and it is a dynamical field. Thus at
early times we have from Eq.~(\ref{JGRG20}) $w_\mathrm{eff}=-1$,
since $w=-1$. On the other hand, after the axion starts to
oscillate, the axion always scales as $\phi(t)\sim a^{-3/2}$, as
it can be seen from Eq.~(\ref{solutionaxionradandaftpaper2}). It
therefore gives an average EoS $\langle w \rangle =0$, for all
cosmic times $m_a\geq H$.

In principle, the radiation domination era solely can be used to
constrain the parameter $\delta$ appearing in the non-minimal
coupling $h(\phi )$. Let us elaborate on this issue, because this
point can be tricky, since during the early-time era, the total
energy density of the cosmological system is determined by the
$R^2$ gravity, due to the fact that the axion is frozen in its
vacuum expectation value, and during the matter domination era,
the axion potential $V\sim \phi^2$ dominates the evolution,
yielding $a(t)\sim t^{2/3}$. Thus the second part of Eq.~(\ref{M8})
which determines the parameter $h_0$ as a function of
the parameter $m$ which is the power of the $F(R)$ gravity, no
longer holds true. The $F(R)$ gravity-curvature related terms do
not control the matter domination era. However, at late times, the
term $h(\phi)G(R)$ controls the evolution, and it actually yields
a asymptotically de Sitter evolution, asymptotic referring to
large cosmic times \cite{Odintsov:2019evb}. Thus the evolution is
not of power-law type at late-times. During the matter domination
era, the potential dominates the evolution, thus since $\phi^2\sim
a^{-3}$ the solution $V\sim\phi^2\sim a^{-3}\sim t^{-2}$, when
equated to the total energy density evolution it yields
$-3h_0=-2$, hence $h_0=2/3$ which is the expected behavior during
the matter domination. Hence the only era for which both the
potential and the term $h(\phi)G(R)$ might be equally dominant is
the radiation domination era, and the combined action of the two
yields a power law scale factor of the form $a\sim t^{1/2}$. If
one of the two was dominant, we would either have a matter
domination era, or an asymptotic de Sitter solution as was evinced
in Ref.~\cite{Odintsov:2019evb} and we previously demonstrated.
This is a speculation though and strong numerical analysis is
needed, but let us assume that this is the case. In such a case,
$h_0=1/2$ during the radiation domination, and hence $\rho\sim
t^{-3/2}$ so by equating this to $h(\phi)G(R)\sim \phi^{-\delta}
R^\gamma \propto \phi^{-\delta} t^{-2\gamma}$ and solving with
respect to $\gamma$, we get,
\begin{equation}\label{para}
\gamma=\frac{1}{4} (-3) (\delta -1)\, .
\end{equation}
Since $0<\gamma<0.75$, Eq.~(\ref{para}) imposes a constraint on
the values of the parameter $\delta$, which is the following,
\begin{equation}\label{dconstraint}
0<\delta <1\, ,
\end{equation}
which is obtained by the assumption that during the radiation
domination era, both the potential and the non-minimal coupling
term are dominant, however this a crude estimate and rigid
numerical analysis is needed to effectively quantify the dynamics
of this era, so we refrain from going into further details. Let us
now consider the energy conditions, to efficiently examine if
these hold true for the model~(\ref{mainaction}). In the standard
Einstein gravity, the energy conditions are given for $\rho$ and
$p$ in the FRW universe,
\begin{eqnarray}
\label{phtm11}
&\circ &\ \mbox{NEC:} \ \rho + p \geq 0\\
\label{phtm8}
&\circ &\ \mbox{WEC:} \ \rho\geq 0 \ \mbox{and}\ \rho
+ p \geq 0 \\
\label{phtm9}
&\circ &\ \mbox{SEC:} \ \rho + 3 p \geq 0\
\mbox{and}\
\rho + p \geq 0\\
\label{phtm10}
&\circ &\ \mbox{DEC:}\ \rho\geq 0 \ \mbox{and}\
\rho \pm p \geq 0
\end{eqnarray}
The above conditions are rewritten in terms of the EoS parameter $w$ if we assume
$\rho\geq 0$.
\begin{eqnarray}
\label{phtm11B}
&\circ &\ \mbox{NEC, WEC:} \ w \geq -1\\
\label{phtm9B}
&\circ &\ \mbox{SEC:} \ 3w \geq - 1\
\mbox{and}\ w \geq -1\\
\label{phtm10B}
&\circ &\ \mbox{DEC:}\ 1\pm w \geq 0
\end{eqnarray}
Even for the models in this paper, we may require,
\begin{eqnarray}
\label{phtm11C}
&\circ &\ \mbox{NEC, WEC:} \ w_\mathrm{eff} \geq -1\\
\label{phtm9C}
&\circ &\ \mbox{SEC:} \ 3w_\mathrm{eff} \geq - 1\
\mbox{and}\ w_\mathrm{eff} \geq -1\\
\label{phtm10C}
&\circ &\ \mbox{DEC:}\ 1\pm w_\mathrm{eff} \geq 0
\end{eqnarray}
It is obvious that the matter content of the model~(\ref{mainaction})
satisfies the energy conditions quantified in
terms of $w$, since for early times $w=-1$, while for all later
times, $w=0$. With regard to the total EoS parameter
$w_\mathrm{eff}$, the values it takes during all eras are as
follows,
\begin{align}\label{weffvalues}
& w_\mathrm{eff}\sim -1,\,\,\,\,\mathrm{inflation}\, , \\ \notag &
w_\mathrm{eff}\sim 1/3,\,\,(h_0=1/2),\,\,\,\,\mathrm{radiation}\,\,\,\mathrm{domination}\, , \\
\notag & w_\mathrm{eff}\sim 0,\,\,\,\,\mathrm{matter} \,\,\,\mathrm{domination}\\
\notag & w_\mathrm{eff}=-1,\,\,\,\mathrm{late}
\,\,\,\mathrm{times}\, ,
\end{align}
so all the energy conditions quantified in terms of
$w_\mathrm{eff}$ are satisfied.

\section{Conclusions}

In this paper we investigated several theoretical and
phenomenological issues related to axion-$F(R)$ gravity in the
presence of a non-minimal coupling of the axion scalar to the
scalar curvature. Particularly, we investigated the ghost-free
conditions and the Einstein frame implications of the model in
general. Also we demonstrated how a general cosmological solution
may be realized by the scalar-$F(R)$ gravity, by using the
equations of motion in the Jordan frame as the main constituents
of a reconstruction method. Also, we examined the gravitational
waves of the scalar-$F(R)$ gravity theory, and we demonstrated
that the gravitational wave modes have the same propagation speed
as in $F(R)$ gravity, but in some cases enhancement or dissipation
may occur, an effect similar to that of a propagation of a gravity
wave in a viscous fluid. Finally, we performed a general study of
the conserved quantities and of the energy momentum tensor, and we
investigated whether  the energy conditions are satisfied. Also we
discussed the phenomenological constraints that are imposed on the
non-minimal coupling of the axion to the scalar curvature during
the radiation domination era. A future perspective of this work is
to extend the gravitational wave study in terms of string theory
originating Chern-Simons gravity, as in
Ref.~\cite{Odintsov:2019mlf}. The existence of polarization asymmetry
in the gravitational wave spectrum, may indicate that a potential
candidate theory that can harbor such effects is an axion-$F(R)$
gravity with axion Chern-Simons terms.

\section*{Acknowledgments}

This work is supported by MINECO (Spain), FIS2016-76363-P, and by
project 2017 SGR247 (AGAUR, Catalonia) (S.D.O). This work is also
supported by MEXT KAKENHI Grant-in-Aid for Scientific Research on
Innovative Areas ``Cosmic Acceleration'' No. 15H05890 (S.N.) and
the JSPS Grant-in-Aid for Scientific Research (C) No. 18K03615
(S.N.).

\end{document}